\documentclass[runningheads]{llncs}
\usepackage[T1]{fontenc}
\usepackage{hyperref}
\usepackage{graphicx}
\usepackage{color}

\begin{document}

\title{An Architecture for Privacy-Preserving Telemetry Scheme}
\titlerunning{Private Telemetry System}

\author{Kenneth Odoh \orcidID{0000-0003-4892-4301}}
\institute{\email{kenneth.odoh@gmail.com}\\
\url{https://kenluck2001.github.io}}

\authorrunning{K.Odoh}

\maketitle              

\begin{abstract}

We present a privacy-preserving telemetry aggregation scheme. Our underlying frequency estimation routine works within the framework of differential privacy. The design philosophy follows a client-server architecture. Furthermore, the system uses a local differential privacy scheme where data gets randomized on the client before submitting the request to the resource server. This scheme allows for data analysis on de-identified data by carefully adding noise to prevent re-identification attacks, thereby facilitating public data release without compromising the identifiability of the individual record. This work further enhances privacy guarantees by leveraging Oblivious HTTP (OHTTP) to achieve increased privacy protection for data in transit that addresses pre-existing privacy vulnerabilities in raw HTTP. We provide an implementation that focuses on frequency estimation with a histogram of a known dictionary. Our resulting formulation based on OHTTP has provided stricter privacy safeguards when compared to trusting an organization to manually delete identifying information from the client's request in the ingestor as deployed in reference work~\cite{apple2017}.
\\\textbf{Source code} available at \url{https://github.com/kenluck2001/miscellaneous/tree/master/src/Privacy-Preserving-Telemetry}

\end{abstract}

\section{Introduction}

Understanding the usage patterns of deployed devices can provide insight into improving customer experience from an organizational perspective. De facto attempts to obtain user data can increase privacy risks. Coincidentally, there is a market for trading customer data to facilitate precise advertisement targeting. Acxiom~\footnote{\url{https://www.acxiom.com/}} is one of the world's largest data brokers that harvests data from nearly a billion users worldwide. A privacy attack can result from the actions of malicious actors who act surreptitiously and pursue goals that are inconsistent with those of the users. Increasing financial motives for abusing users' data have motivated our research into novel privacy-enhancing mechanisms to provide privacy by design.

Internet standards~\footnote{\url{https://www.ietf.org/standards/}} led by a consortium of technology firms, university researchers, hobbyists, and others have a higher chance of building higher quality internet protocols due to the increased scrutiny of multiple industrial partners. Similarly, these setups are analogous to peer reviews in that they help researchers improve their engineering thinking through the critical evaluations of their work. In the same way, it is prudent to build industrial systems utilizing the privacy guarantees afforded by the Oblivious HTTP protocol rather than blindly trusting an Apple internal aggregator service. Given how recent news has demonstrated the prevalence of blatant privacy abuse of user data in the industry. Hence, we have taken this approach in this manuscript.

Differential privacy (DP) is a structured mathematical framework that supports principled reasoning about privacy loss in a database. Randomization happens by adding calibrated noise to the original data to prevent reverse-engineering the original value of the randomized data, thereby providing privacy protection. DP protects sensitive data while maintaining a trade-off between added noise and expected utility. As a result, DP has increased the adoption of privacy-preserving data mining tasks that facilitate public data release without compromising individual privacy. The rigorous nature of the DP mechanism makes it ideal for satisfying evolving privacy regulations.

Gathering telemetry is a necessary prerequisite for several data analytics tasks. Our work has adopted differential privacy as a standard for guaranteeing privacy protection. This work extends the privacy guarantee of the system built by Apple~\cite{apple2017}. Their work~\cite{apple2017} requires trust that the ingestor will not abuse client-identifying data. This expectation of trust is unrealistic as monetary benefits arise from potential trading on customer data. The unanswered question is how to improve privacy on this system without resorting to onion routing~\cite{navalresearch2002}? TOR utilizes flooding in its operation and incurs unacceptable overhead that can impact scalability.
In contrast, systems such as Prio~\cite{Henry2017}, DPrio~\cite{Keeler2023}, and Prio+~\cite{Addanki2022} may provide higher privacy guarantees due to their incorporation of multiparty-based secret sharing. On the contrary, several applications require reasonable privacy guarantees with minimal setup costs, which is the premise of our manuscript.

Our thesis focuses on enhancing the privacy of our telemetry scheme based on Oblivious HTTP~\cite{thomson2021} with significant simplification. We seek to understand common user patterns across devices by generating snapshot readings for summary device health or other information. Hence, we have developed a privacy-preserving telemetry system that uses local differential privacy, where data gets randomized on the client before submitting the request to the resource server. Therefore, it delivers a higher degree of privacy guarantee when compared to the central differential private scheme. The paper is structured as follows: a summary of contributions in Section~\ref{contributions}, a literature review of previous works in Section~\ref{related-work}, a brief explanation of differential privacy in Section~\ref{diff-priv}, Oblivious HTTP in Section~\ref{ob-http}, an overview of our base implementation in Section~\ref{system-overview}, a discussion of the merits of our solution in Section~\ref{discuss}. We have demonstrated the usefulness of our architecture with a case study and several experiments in Section~\ref{case-study}. Finally, we present limitations, future work, and conclusions in Section~\ref{Future work} and Section~\ref{conclude}.

\section{Contributions}
\label{contributions}

Our contributions are summarized as follows:

\begin{itemize}
\item We provide an implementation focused on frequency estimation with a histogram of known dictionary words. However, the reference work~\cite{apple2017} has a known limitation where we trust the ingestor will not cooperate with bad actors that may abuse the customer's privacy.
\item HTTP does not provide privacy by design. The quest for rigorous privacy protection has motivated us to build our privacy scheme based on Oblivious HTTP, which fixes many privacy vulnerabilities in HTTP.  
\item We demonstrate a conceptual framework for enhancing privacy protection using a standardized internet protocol. Therefore, we no longer need to trust the ingestor will not abuse client identifiers (such as IP addresses or session data). This setup~\cite{apple2017} results in a weaker notion as it is difficult to audit whether the required deletion happened.

\end{itemize}

\section{Related Work}
\label{related-work}

Private telemetry is very interesting to all service providers, as seen in all major browsers and operating systems. Procho~\cite{Bittau2017} introduced the Encode, Shuffle, Analyze(ESA) framework widely used in telemetry, error reporting, and continuous monitoring. STAR~\cite{Davidson2022} is a data aggregation system that enforces k-anonymity based on well-known cryptographic primitives. Privacy leaks from this scheme can impact users' confidence. Similarly, several deployed telemetry systems exist in the industry, but most lack privacy-preserving characteristics. For example, Facebook created a system named PCAT~\cite{ZhouTel2022} to continuously monitor production assets and offer support for change detection, alerting, monitoring, and diagnostics. As a result, when this telemetry scheme gets deployed beyond the sandboxed production environment to real-user devices on the edge. Privacy leaks from this scheme can impact users' confidence. Subsequently, privacy-preserving data mining methods have evolved from theoretical abstractions to solving real-world applications.

Differential privacy~\cite{dwork2017} (DP) is a robust method for quantifying how privacy degrades under frequent adversarial evaluation of database records. DP can work in local or central settings where a local DP scheme has higher privacy guarantees. There are examples of public-facing industrial DP deployment in local settings such as RAPPOR~\cite{fanti2016} and central settings in PINQ~\cite{mcSherry2009}. Furthermore, alternative notions of privacy-enhancing technology include Verifiable Distributed Aggregation Functions (VDAFs)~\cite{patton2021} and TOR~\cite{navalresearch2002} can provide more privacy guarantees at a higher cost than Oblivious HTTP~\cite{thomson2021}.

Several lines of work utilize sketch-based algorithms for network monitoring because of their efficient approximate count estimation as follows: OctoSketch~\cite{ZhangOct2024}, TrustSketch~\cite{ChengTrust2022}, and HeteroSketch~\cite{AgarwalSketch2022}. Hence, we have adopted sketch-based estimation as seen in (Algorithms 1, 3, 5, and 7) of our reference paper~\cite{apple2017}. Several privacy-preserving analytics processing engines exist to support downstream data analysis. One such scheme is PRIVAPPROX~\cite{Quoc2017} utilizing a zero-knowledge proof construction to provide higher privacy guarantees than differential privacy. POPSTAR~\cite{Li2024} uses oblivious PRF and polynomial commitment for privacy-preserving aggregation schemes.

One such case is the privacy-aware deployment at Apple~\cite{apple2017} to capture insights into crashes (and other events) from a collection of phones using well-known security policies and differential privacy. Through our work, we propose privacy-preserving frequency estimation without trusting that the ingestor will delete client-identifying information without a persistent audit. We have eliminated the trust by providing a simplified implementation with extended privacy guarantees by adopting Oblivious HTTP~\cite{thomson2021} to increase clients' privacy assurance.

\section{Differential Privacy}
\label{diff-priv}

Differential Privacy (DP) is a privacy-enhancing technology that allows for data analysis on de-identified data by carefully adding noise to prevent re-identification attacks, thereby facilitating public release without impacting the privacy of the individual record.

\textbf{Definition 1}: (Differential Privacy) Following Definition 7 of~\cite{dwork2017}, for each pair of the data record $D$ and $D^{\prime}$, noise, $\epsilon$, and a randomizer, $\mathcal{M}$ satisfies $\mathrm{P}(\mathcal{M}(D) \in \mathcal{O}) \leq e^\epsilon \mathrm{P}\left(\mathcal{M}\left(D^{\prime}\right) \in \mathcal{O}\right)$

When $\epsilon \approx $0, we attain higher privacy guarantees with more similarities across the data set. Note, when $\epsilon = $0, at that point, perfect secrecy is achieved by limiting the ability to perform statistical analysis. When $\epsilon = \infty$, we have a blatantly non-private mechanism. Therefore, we aim to achieve reasonable privacy within an appropriate budget.

\section{Oblivious HTTP}
\label{ob-http}

Oblivious HTTP~\cite{thomson2021} (OHTTP) is an encapsulated abstraction built on top of HTTP to address inherent privacy risks within communicating peers. There are some industrial deployments of privacy-enhancing protocols such as iCloud private relay~\cite{rustam2022} and Flo period tracker app~\cite{flo2022}.

This OHTTP protocol allows exchanging encrypted messages where the server cannot link the request to the client. This setup eliminates the risk of leaking client information when communicating. For example, exposing an IP address can uncover an individual, as it is an identifier linked to a physical node (client), or reveal information about an IoT device in a house. We can mitigate privacy leaks between communicating parties (client-server architecture) using HTTP. OHTTP operates by using a proxy (relay server) to send the request between the client and server by adopting this level of indirection to prevent request linkability.

We have added a simplification where we removed the gateway and instead used a 3-party system (client, relay server, resource server) instead of the 4-party system (client, relay server, gateway server, resource server) as defined in the standard~\cite{thomson2021}. 

\begin{figure}[h]
    \centering
    \includegraphics[scale=0.55]{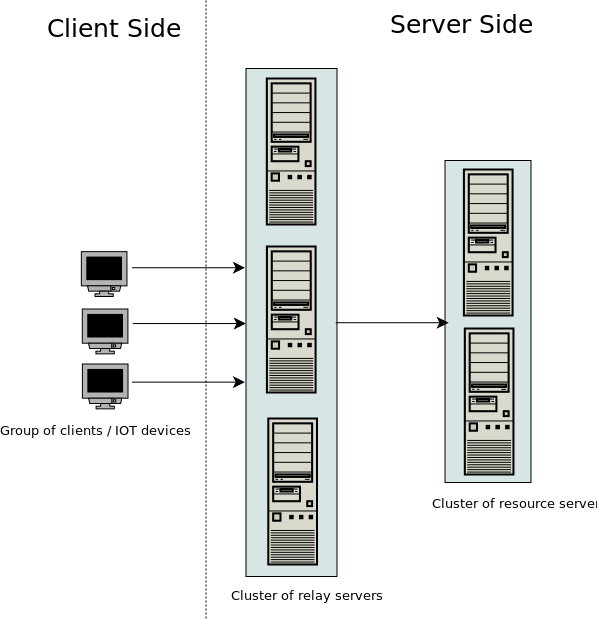}
    \caption{Simplified Oblivious HTTP}
    \label{fig:ohttp}
\end{figure}

The description of our simplified Oblivious HTTP protocol considering only the request flow (response flow got omitted because our requirement is unidirectional) is as follows, as shown in Figure~\ref{fig:ohttp}:
\begin{itemize}
\item The Client creates an encrypted message using the (public key of the resource server) and forwards it to the relay server.
\item The relay server forwards the encrypted message to the resource server. It is a requirement that the relay server cannot read the message, as it does not have the required key to decrypt the message on the relay server. The relay transfers information without knowing the message content.
\item The resource server can decrypt the message using its private key.
\end{itemize}

\section{System Overview}
\label{system-overview}

\begin{figure}[h]
	\centering
	\includegraphics[scale=0.65]{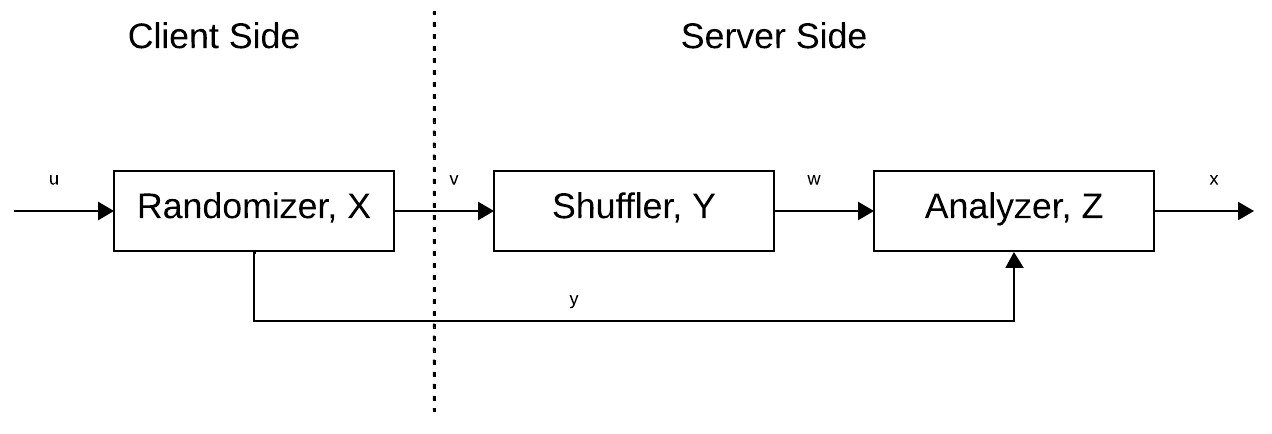}
	\caption{DP Architecture~\cite{Bittau2017}}
	\label{fig:architecture}
\end{figure}

This work demonstrates privacy-aware frequency aggregation of event telemetry. Our implementation focuses on frequency estimation of events where we compute a histogram from a known word distribution. This formulation allows counting the frequency of a term from a known dictionary of terms. Furthermore, we have provided an aggregate of known terms as event identifiers. Apple deployment follows an equivalent naming convention with substitute names as described in paper~\cite{balcer2019} as shown in Figure~\ref{fig:architecture}.

\begin{figure}[h]
	\centering
	\includegraphics[scale=0.19]{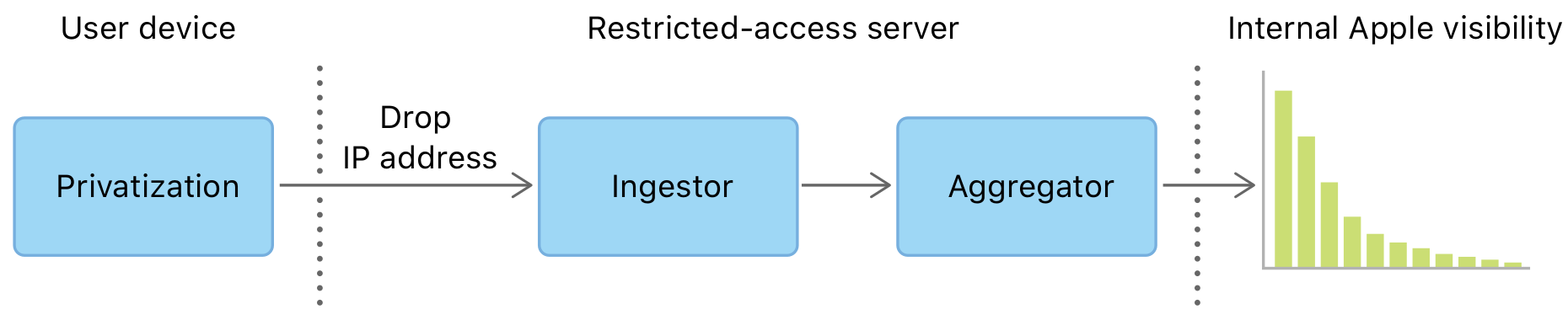}
	\caption{Apple DP system~\cite{apple2017}}
	\label{fig:applesystem}
\end{figure}

Where $u, v, w, x, y$ are variables depicting the life cycle of the data as it transits different stages through the pipeline as given the input data, $u$ and resulting system output data, $y$, and transformation functions $X, Y, Z$ are randomizer, shuffler, and analyzer respectively. The shuffler is optional based on the use case. This data processing pipeline has the following phases: randomizer (privatization), shuffler (ingestor), and analyzer (aggregator) as shown in Figure~\ref{fig:applesystem}. From the perspective of a single user, they randomize the data in their device and send it to the ingestors, where identifying information is removed, and the resulting data gets forwarded to the analyzer, where aggregate statistics get computed.

We have implemented the following algorithms from the paper~\cite{apple2017} that include $A_{client-HCMS}$ in Algorithm 5 of~\cite{apple2017}, Hadamard count mean sketch HCMS in Algorithm 6 of~\cite{apple2017}, $A_{server}$ in Algorithm 4 of~\cite{apple2017}, and computing the Sketch matrix known as Sketch-HCMS in Algorithm 7 of~\cite{apple2017}. We extend previous work~\cite{apple2017} by utilizing a proven privacy mechanism in Oblivious HTTP~\cite{thomson2021}. We can replace the ingestor shown in Figure~\ref{fig:applesystem} with a relay server in the setup of Oblivious HTTP. The relay server does not know anything about the requests forwarded through it. Hence, our scheme based on OHTTP has provided a stricter privacy safeguard when compared to trusting an organization to manually delete identifying information from the client's request in the ingestor.

The DP algorithm has a server and client mechanism shown in Figure~\ref{fig:ohttp}, where the client-side algorithm is a locally differentially private scheme where the client randomized each data instance before sending transformed results to the server. Consequently, the server provides a near-precise count of events where aggregated results can handle customers' information-seeking needs. Data transfer between the client and the server can impact communication costs. Calibrated noise added to the client during the privatization phase can dictate the amount of privacy afforded and achievable accuracy at the analyzer stage. Therefore, we can achieve a trade-off between privacy, communication cost, and computation accuracy. The server-side algorithm averages the count for $m$ number of hash functions. Similarly, the hash function should be a set of $m$ instances of 3-wise independent hash functions where $m$ is the number of hash functions. 

\textbf{$k$-wise independent hash function}~\footnote{\url{https://www.cs.purdue.edu/homes/hmaji/teaching/Fall\%202017/lectures/12.pdf}}: $k$-wise independent is satisfied for a set of discrete random variables $X_1, \ldots, X_n$ given that for any set $I \subseteq\{1, \ldots, n\}$ with $|I| \leq k$ and any values $x_i$ as shown in Equation~\ref{eqn:kwiseindep}.

\begin{equation}
\label{eqn:kwiseindep}
\textbf{Pr}\left[\wedge_{i \in I} X_i=x_i\right]=\prod_{i \in I} \textbf{Pr}\left[X_i=x_i\right] .
\end{equation}

Following Equation~\ref{eqn:kwiseindep}, $k$-wise independence is satisfied if we can choose a function from a hash family with a guarantee that any $k$ keys are independent random variables. One example of a $k$-wise independent hash function follows the polynomial structure as shown in Equation~\ref{eqn:kwise}.

\begin{equation}
\label{eqn:kwise}
H(x) = a_{0} + a_{1}x + a_{2}x^2 + ... + a_{k-1}x^{k-1}
\end{equation}

Where $H(x$) is a polynomial of degree $\le k$

\section{Case Study}
\label{case-study}

Sports tracking apps are vulnerable to adversary eavesdropping and capturing user activity to target individuals as recent news has shown~\footnote{\url{https://www.cnn.com/2023/07/11/europe/russian-submarine-commander-killed-krasnador-intl/index.html}}. Following the prevalence of privacy abuse, we present a case study on one of our users named "Jane" who is a sports enthusiast and uses our fictitious generic sports tracking app~\footnote{\url{https://www.bbc.com/news/av/technology-24379432}}, who wears a tracking device. We have demonstrated our privacy-preserving telemetry architecture with sport tracking. Our scheme supports a client-server architecture. The client side is the device tracker worn on the person, while the server component aggregates the data.

Our implemented scheme protects users' privacy as data gets randomized on the clients before transferring the same data to the server. As a result, we can achieve local differential privacy in our implementation. By utilizing this architecture, there is less chance of privacy risk, as an attacker can intercept the scrambled data on transmission to the server, and the adversary becomes incapable of reverse-engineering to uncover the original data from the randomized data. The device tracker uses sensors such as a GPS locator, gyroscope, accelerometer, and others to categorize the activities of users into telemetry events that include: "walking", "running", and "sleeping".

As part of our evaluation, we have arbitrarily randomly sampled these events ("walking", "running", and "sleeping") using the given probabilities ($\frac{3}{5}, \frac{3}{10}, \frac{1}{10}$). The sampled data is a list of snapshots of original data (telemetry events) on the client device and gets transformed using the privatization algorithm. The de-identified scrambled data is then transferred to the aggregator (relay server) and then to the analyzer, where aggregate statistics are estimated. We can claim that our telemetry scheme works as expected if the distribution of the events after the analyzer stage matches the data distribution of the original sampled events before randomization.

Jane always wears a device tracker to monitor her activities for health reasons. Let us define two concepts used in our discussion.

\begin{itemize}

\item Original data proportion: This measure is the ratio of the occurrence of telemetry events captured in the data before randomization on the clients. For example, if the data has the following events as follows: 10 "walking", 5 "running", and 5 "sleeping", then the resulting probabilities are ($\frac{1}{2}, \frac{1}{4}, \frac{1}{4}$) for ("walking", "running", and "sleeping") respectively. 

\item Randomized data proportion: This measure is the ratio of the occurrence of telemetry events captured in the data (after randomization). For example, if the data has the following events as follows: 10 "walking", 5 "running", and 5 "sleeping", then the resulting probabilities are ($\frac{1}{2}, \frac{1}{4}, \frac{1}{4}$) for ("walking", "running", and "sleeping"). 

\end{itemize}

We have provided two experiments to demonstrate the usefulness of our implementation as shown in SubSections~\ref{distribution-mismatch},~\ref{noise-randomized-data-distribution}, and~\ref{noise-interpret-data-distribution}. The y-axis is the count of telemetry events after randomization in Figures~\ref{figure:distribution-mismatch} and~\ref{figure:noise-randomized-data-distribution}.

\subsection{Distributional mismatch between analyzer output and original data (input)}
\label{distribution-mismatch}

\begin{figure}[h]
	\centering
	\includegraphics[scale=0.42]{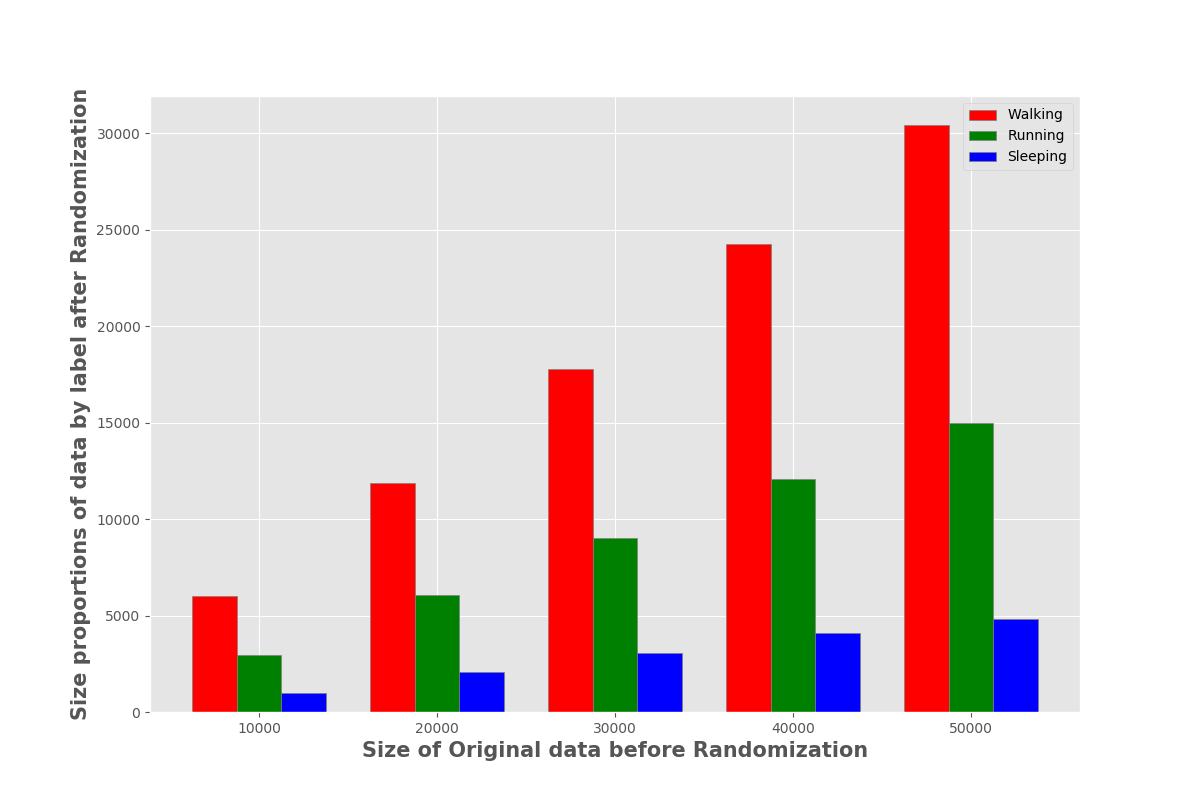}
	\caption{Impact of original data size (before randomization) on proportion of randomized data}
    \label{figure:distribution-mismatch}
\end{figure}

We have designed an experiment to understand how the randomized data proportion of telemetry events varies as the data size increases. The original data probability is kept constant for random sampling as ($\frac{3}{5}, \frac{3}{10}, \frac{1}{10}$) for ("walking", "running", and "sleeping") respectively as shown in Figure~\ref{figure:distribution-mismatch} with the noise, $\epsilon = 4$, and the data get increased to observe the influence on the randomized data proportions. The x-axis is the original telemetry count before privatization in Figure~\ref {figure:distribution-mismatch} and the combined telemetry count (y-axis) due to the approximate nature of the sketch-based frequency algorithm.

Similarly, we can see from Figure~\ref{figure:distribution-mismatch} that the randomized data proportion of events (after privatization) does not significantly change. This phenomenon implies that the privatization algorithm does not impact its utility at the set noise level. Our implementation shows that the frequency counting estimate is robust and preserves the distribution of the original data (before randomization).

\subsection{Noise level impact on randomized data distribution}
\label{noise-randomized-data-distribution}

\begin{figure}[h]
	\centering
	\includegraphics[scale=0.42]{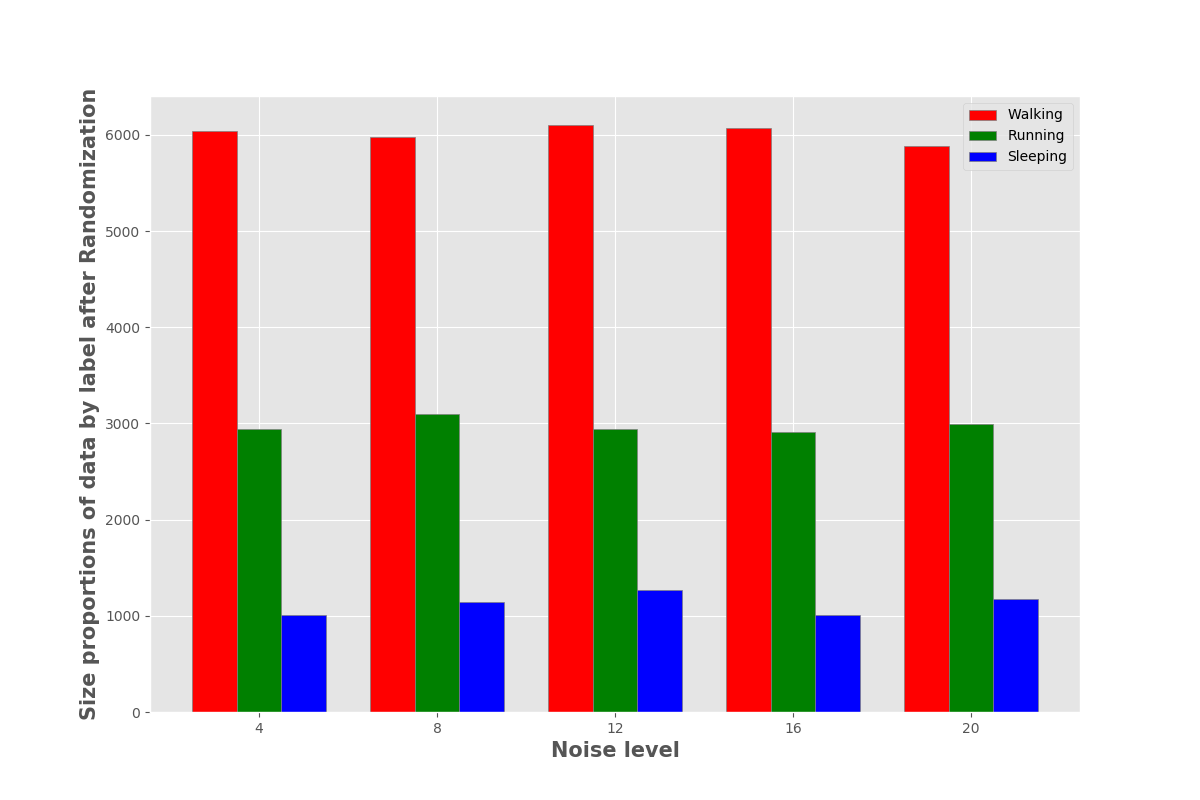}
	\caption{Impact of noise levels on proportions of randomized data}
    \label{figure:noise-randomized-data-distribution}
\end{figure}

The experiment demonstrates how the proportions of randomized data change with increasing noise levels shown in Figure~\ref{figure:noise-randomized-data-distribution}. The x-axis is the telemetry count before privatization in Figure~\ref{figure:noise-randomized-data-distribution}.

Furthermore, we can observe from Figure~\ref{figure:noise-randomized-data-distribution} how increasing the noise, $\epsilon$, during randomization changes the proportion of randomized data.

\subsection{Interpreting experimental results}
\label{noise-interpret-data-distribution}

We have derived insights from the experimental results visualized in Figures~\ref{figure:distribution-mismatch} and~\ref{figure:noise-randomized-data-distribution}. The results show that the accuracy of the total count of original telemetry events can vary with the estimate obtained from the analyzer after applying the sketch count algorithm for frequency estimation on the telemetry event. The estimated count is an approximation of the actual telemetry event count. However, the total count of a class of telemetry elements from the sketch-based frequency algorithm in this work ensures that the data proportions are relatively fixed even if the data size keeps increasing. We also observe that the proportion of randomized events is relatively constant in the face of increasing noise levels, so the predefined noise in our sketch-based frequency estimation within reasonable bounds does not impact the counting process. This finding implies that we obtain an approximate count of telemetry events. It captures the trend in the data as the original data distribution gets relatively unchanged from the output distribution of the analyzer phase within reasonable bounds.

\section{Discussion}
\label{discuss}

The system architecture demonstrated in this manuscript supports a variety of use cases. First, we can provide a platform for analyzing the sports activities of a group of users while restricting the identifiability of a single user without hindering the applicability of understanding the collective actions of individuals under observation. Second, we can organize each user's sporting events into groups and relax our privacy definition, where each group is linkable to the individual, and the randomized events in the group result in uncertainty in each event. For example, after randomization, a user "sleeping" may be confused with the same user "running" at a given time. As a result, individuals can analyze their sporting activities over an extended time horizon with a reasonable data size. The user gleans information about aggregated sporting activities. Furthermore, the successful deployment of differential privacy-based systems requires a principled way to determine the noise, $\epsilon$, with minimal influence on the system's utility. As a result, several lines of work~\cite{Laud2019},~\cite{KenGroup2024} have focused on estimating the optimal noise magnitude, $\epsilon$.

Setting up OHTTP requires a set of precautions to prevent privacy violations. OHTTP mandates that each request be stateless to avoid correlations between requests that can impact privacy and uncover the identity of the connecting client. OHTTP provides privacy, given that the relay and the resource server do not form a collusion ring. Our approach favors forwarding over flooding. As a result, we favor the forwarding scheme in OHTTP instead of TOR~\cite{ManiWJJS18}. Relaying (forwarding by proxy) can likely impact latency. However, OHTTP fits nicely within the pre-existing internet infrastructure. We built our infrastructure on the foundation of distributed computing principles, including replication and failover, to provide high availability for the relay server. Hence, the existing fault-tolerant setup is sufficient for our unidirectional scheme.

We have used two layers of security, where the channel is secured using public key cryptography as part of the OHTTP protocol, and the data itself gets transformed utilizing differential privacy as part of the telemetry scheme based on the paper~\cite{apple2017}. Denial-of-service and replay attacks are other security challenges when using OHTTP. This protocol can prevent denial-of-service attacks by being rate-limited. An attacker can stage a replay attack by positioning a rogue server to monitor packet traffic. Subsequently, OHTTP has a built-in method for mitigating replay attacks~\footnote{\url{https://www.rfc-editor.org/rfc/rfc8446\#section-8}}.

Let us consider the implementation intricacies of our DP implementation. Our approach to creating a family of hash functions, $H(x)$ shown in Equation~\ref{eqn:kwise} was to generate a random matrix and extract the parameter for each hash function using vectors obtained row-by-row or column-by-column based on matrix dimensions. This setup provides an advantage to having a set of shared hash family functions for sampling the hash functions for both the client and the server. Another approach is to create a family of hash functions on the server and send them to the client. This scenario may be undesirable if we consider communication costs. A compromise solution may bias our frequency count estimates by utilizing different hash function families for clients and servers with similar distributions. As a result, we avoid sending huge matrices over the network. Eventually, a better optimization is to adopt an identical random seed with the same Pseudorandom Generator~\footnote{\url{https://en.wikipedia.org/wiki/Pseudorandom_generator}} (PRG) on the client and server, thereby enabling local sampling from the same hash family distribution without communication costs. Hadamard transforms help to reduce the variance of estimates at the analyzer (aggregator) stage as shown in Figures~\ref{fig:architecture} and~\ref{fig:applesystem}. The Hadamard count sketch algorithm is an optimized variant utilizing a dense vector instead of a sparse matrix. We observed that the quality of the solution depends on the choice of a hash function. Therefore, we created a custom hash function with appropriate statistical properties for ASCII~\footnote{\url{https://en.wikipedia.org/wiki/ASCII}} strings.

\section{Limitations and Future Work}
\label{Future work}

Our work has a limitation due to using a family of hash functions that support only ASCII strings, thereby restricting our ability to handle events in wide-character languages requiring more than 8 bits to represent a character. Furthermore, we can improve our work by categorizing client devices by utilizing a set of gateway servers in our Oblivious HTTP flow to support the logical grouping of requests. Also, the relay server can strengthen privacy protection by using anycast~\cite{sommese2020} address on a cluster of relays by purposefully increasing uncertainties linking a client's request to a particular relay server. It is imperative to ensure that the relay servers are not under any big tech firm to prevent compromise.

\section{Conclusion}
\label{conclude}

We have extended the privacy-preserving frequency of event telemetry~\cite{apple2017} with oblivious HTTP. Furthermore, we have provided a working implementation of the privacy-preserving architecture with several significant improvements. Our work would facilitate a privacy-aware telemetry system for obtaining internet measurements (or other measures) while providing privacy protection.

\bibliographystyle{splncs04} 
\bibliography{sample-sigconf}

\end{document}